\documentclass[aip,
 amsmath,amssymb,
 reprint,%
]{revtex4-1}

\usepackage{graphicx}% Include figure files
\usepackage{dcolumn}% Align table columns on decimal point
\usepackage{bm}% bold math
\usepackage{booktabs}
\usepackage{amssymb} % for \checkmark
\usepackage{siunitx}
\usepackage[noend]{algorithmic}
\usepackage{algorithm}
\usepackage[skins]{tcolorbox}%colorbox
\usepackage{setspace}
\usepackage{blkarray}
\usepackage{tabularx}
\usepackage{xcolor}
\usepackage{soul}
\usepackage{systeme}
\usepackage{scalerel} % To change size things in math environment \scaleto{...}{1pt}
\usepackage{placeins}
\makeatletter
\def\maketag@@@#1{\hbox{\m@th\normalfont\normalsize#1}}
\makeatother

\newcommand{\tr}{\text{tr}}

\usepackage{booktabs}
\usepackage{hyperref}
\usepackage[capitalize]{cleveref}

\usepackage[utf8]{inputenc}

\usepackage[lining,semibold]{libertine} % a bit 
\usepackage{amsthm}
\usepackage[libertine, cmintegrals, bigdelims, vvarbb]{newtxmath}

\usepackage{amsmath}
\usepackage{amsfonts}
\usepackage{mathrsfs}
\usepackage{gensymb}
\usepackage{bbm}
\usepackage{dsfont}

\usepackage{pgfplots}
\pgfplotsset{compat=1.7}

\usepgfplotslibrary{colormaps}

\usepackage{chemformula}
\usepackage[caption=false]{subfig}

\usepackage{soul}
\usepackage{xcolor}
\definecolor{webgreen}{rgb}{0,.5,0}
\definecolor{webbrown}{rgb}{.6,0,0}
\definecolor{grigio}{rgb}{.85,.85,.85} 
\definecolor{RoyalBlue}{rgb}{0.0, 0.14, 0.4}
\definecolor{skyblue1}{rgb}{0.45,0.62,0.81}
\definecolor{skyblue2}{rgb}{0.2,0.39,0.64}
\definecolor{skyblue3}{rgb}{0.13,0.29,0.53}
\definecolor{scarlet1}{rgb}{0.93,0.16,0.16}
\definecolor{scarlet2}{rgb}{0.8,0,0}
\definecolor{scarlet3}{rgb}{0.64,0,0}

\definecolor{g}{gray}{0.50}

\hypersetup{%
    colorlinks=true, linktocpage=true, pdfstartpage=1, pdfstartview=FitV,%
    breaklinks=true, pdfpagemode=UseNone, pageanchor=true, pdfpagemode=UseOutlines,%
    plainpages=false, bookmarksnumbered, bookmarksopen=true, bookmarksopenlevel=1,%
    hypertexnames=true, pdfhighlight=/O,%nesting=true,%frenchlinks,%
    urlcolor=webbrown, linkcolor=RoyalBlue, citecolor=webgreen, %pagecolor=RoyalBlue,%
    pdftitle={},%
    pdfauthor={Timur Aslyamov},%
    pdfsubject={},%
    pdfkeywords={},%
    pdfcreator={pdfLaTeX},%
    pdfproducer={LaTeX REVTeX}%
}

\theoremstyle{definition} 

\usepackage{scalerel} 
\usepackage{etoolbox}
\begin{document}

\preprint{AIP/123-QED}

\title{Classification of instabilities for the nonideal Brusselator model}
% Force line breaks with \\

\author{Premashis Kumar}
\email{kumarpremashis@gmail.com}
\author{Massimiliano Esposito}%
\email{massimiliano.esposito@uni.lu}
\author{Timur Aslyamov}
\email{timur.aslyamov@uni.lu}
\affiliation{Complex Systems and Statistical Mechanics, Department of Physics and Materials Science,\\ University of Luxembourg, 30 Avenue des Hauts-Fourneaux, L-4362 Esch-sur-Alzette, Luxembourg
}%

\date{\today}

\begin{abstract}
We investigate a nonideal, thermodynamically consistent Brusselator reaction–diffusion (RD) system that explicitly incorporates molecular interactions among species in both the diffusion process and the underlying chemical reaction network. 
Within this framework, we systematically revisit the Cross–Hohenberg classification of instabilities to assess the feasibility and characteristics of the various types of instability arising from the interplay between entropic and energetic contributions. 
Our analysis demonstrates that only type I and type III instabilities (from the Cross–Hohenberg classification) can occur in this system; Energetic contributions do not explicitly generate instabilities, but may implicitly control their occurrence through their influence on the fixed-point (steady-state) concentrations. In cases where instabilities of different types coexist, we show that the resulting patterns are highly sensitive to the relative strengths of the competing instabilities.
\end{abstract}

\maketitle

\section{Introduction}
The foundation for the reaction-diffusion (RD) framework was laid by A. M. Turing in his seminal work on morphogenesis~\cite{turing1952}, delineating a possible mechanism for spontaneous symmetry breaking that yields stationary patterns in two-variable systems. 
Building on this foundation, the Brussels School of Thermodynamics, under the leadership of I. Prigogine elucidated the physicochemical implications and universality of these patterns~\cite{Lefever2018, Prigogine1971, Nicolis1977}. 
The Reaction-Diffusion (RD) framework has garnered considerable scientific attention due to its ability to encompass a wide range of patterns prevalent in biological systems and complex structures~\cite{Murray2002, hairfolliclepattern, Kondo1616, Halatek2018a, brauns2020phase, GolestanianCIPS, dwick1, speck, Agranov_2024, sorkin2025accelerated, sarkar2025mechanochemical, winkler2025active, blom2025dynamic, elliott2025repulsive,goychuk2024self,rasshofer2025capillary}. Nevertheless, the thermodynamic theory of these patterns and structures was initially confined to near-equilibrium regimes. 
Recently, its applicability has been extended to encapsulate far-from-equilibrium scenarios by utilizing insights from stochastic thermodynamics~\cite{Falasco2018a, Avanzini2019a, Avanzini2020a}. 
However, these theoretical frameworks, along with many others~\cite{marconhigh2016, diego2018key, haas2021turing, brauns2020phase,nagayama2025geometric}, leverage the assumption of ideal solutions, where the concentration dynamics is dictated by linear diffusion and mass-action kinetics, with patterns emerging in the presence of open multimolecular reactions.

In contrast, nonideal mixtures characterized by molecular interactions between species can induce patterns even in the absence of chemical reactions, as described by the Cahn-Hilliard theory of spinodal decomposition~\cite{cahn1958free}. 
These reactions can alter the phase separation dynamics, and when driven out of equilibrium, such active systems exhibit a rich array of nontrivial phenomena~\cite{glotzer1995reaction, carati1997chemical,zwicker2017growth, weber2019physics,ZwickerSuppression, LEE2018, tena2021accelerated, ziethen2023nucleation,nakashima2021active,avanzini2024nonequilibrium, sorkin2025accelerated,zwicker2025physics}. 
Nevertheless, to ensure physicochemical validity and thermodynamic consistency, the RD representations of these active systems must incorporate nonideal chemical potentials in their expressions of diffusion and chemical dynamics~\cite{lefever1995comment, carati1997chemical}. 
Although thermodynamically consistent models have been explored in the literature~\cite{weber2019physics,kirschbaum2021controlling,zwicker2022intertwined,bauermann2022energy}, they primarily focus on unimolecular reactions, which exclude spatial instability without energetic interactions. 

Recently, progress has been made by integrating multimolecular reactions and formulating a theory of instabilities that governs spatial organization through the interplay of chemical reactions and molecular interactions~\cite{aslyamov2023nonideal, Miangolarra2023, blanchini2023robust}. %yin2023contractility
Complementing this progress, a thermodynamically consistent approach has been employed to capture the energetic characteristics of steady-state patterns far from equilibrium~\cite{avanzini2024nonequilibrium}, revealing the fundamental link between the topology of the chemical reaction network and the spatial distribution of the entropy production rates arising from both diffusion and chemical reactions. However, instabilities that generate patterns in the nonideal RD framework have not yet been adequately investigated through the lens of pattern classification, unlike their ideal counterparts, where the Cross-Hohenberg classification scheme~\cite{Cross1993, cross2009pattern} provides a well-established basis for categorizing stationary and oscillatory instabilities based on the behavior of the dispersion curve. 

In this study, we use the Cross–Hohenberg framework to classify patterns in a paradigmatic nonideal RD system, the Brusellator. 
Although the Cross–Hohenberg classification and extensions of it have been previously reported, most notably in the two-field nonreciprocal Cahn-Hilliard model~\cite{Uwethiele1, Uwethiele2, Uwethiele3}, these studies focused on conserved dynamics. 
In contrast, the nonideal Brusellator RD model considered here involves nonconserved dynamics. 
We analytically characterize the possible instabilities, differentiating between stationary and oscillatory ones, as well as the interplay between the entropic and energetic contributions intrinsic to nonideal chemical systems that give rise to them.     

The structure of the paper is as follows. In~\cref{sys}, we define the description of the nonideal Brusselator RD system. In ~\cref{theo}, we analyze the stability of its homogeneous steady state. Then we classify the various pattern-forming instabilities in \cref{sec:ar}. 
% In~\cref{sr} we present the results of the numerical simulation showing the emergence of different types of patterns. 
Conclusions are drawn in~\cref{dis}.

\section{\label{sys}Nonideal Brusselator Model}
We consider an RD system of two internal species $\{X_1, X_2\}$ and four chemostatted species $\{\ch{A}, \ch{B}, \ch{D}, \ch{E}\}$, which are involved in the weakly reversible version of the Brusselator~\cite{Nicolis1977}
\begin{equation}
\begin{split}
        \text{A} &\ch{ <=>[ $1$ ][ $-1$ ]} X_1 \,,\\
        2X_ 1+X_ 2 &\ch{ <=>[ $2$ ][ $-2$ ]} 3X_ 1\,,\\
       \text{B}+X_ 1 &\ch{ <=>[ $3$ ][ $-3$ ]} X_ 2+\text{D}\,,\\
        X_ 1 &\ch{ <=>[ $4$ ][ $-4$ ]}\text{E}\,.
        \end{split}
        \label{eq:crn-brusselator} 
\end{equation}
In~\cref{eq:crn-brusselator}, the chemostatted species are ideal, constant, and homogeneously distributed in space, while the internal species evolve over time, are distributed in space, and interact with each other. To define the energetics of chemical species, we introduce the nonequilibrium Helmholtz free energy $F[\boldsymbol{x}(t,\boldsymbol{r})]$ considering its equilibrium form, but as a functional of the nonequilibrium concentration fields $\boldsymbol{x}=(x_1,x_2)^\intercal$, where $x_i$ are the concentrations of internal species $X_i$, $\boldsymbol{r}=(r_1,r_2)^\intercal$ are the coordinates of a two-dimensional space and $t$ is time. To incorporate the fluid model, the free energy $F$ can be expressed as the addition of the ideal $F_\text{id}$ and the nonideal $F_\text{nid}$ contributions
\begin{equation}
F[\boldsymbol{x}(t,\boldsymbol{r})]= F_\text{id}[\boldsymbol{x}(t,\boldsymbol{r})] + F_\text{nid}[\boldsymbol{x}(t,\boldsymbol{r})]\,,
\label{eq:free}
\end{equation}
where the ideal term,
\begin{align}
    F_\text{id}[\boldsymbol{x}(t,\boldsymbol{r})]=\sum_i \int d\boldsymbol{r} x_i(t,\boldsymbol{r})(\ln x_i(t,\boldsymbol{r}) - 1)\,,
\end{align} 
is the entropic contribution associated with the ideal solution; and the nonideal term, $F_\text{nid}$, accounts for the molecular (energetic) interactions between the chemical species and can be calculated as
\begin{align}
    F_\text{nid}[\boldsymbol{x}(t,\boldsymbol{r})] = &\sum_{ij}\int d\boldsymbol{r} x_i(t,\boldsymbol{r}) U_{ij} x_j(t,\boldsymbol{r}) \nonumber \\ 
    &+ \sum_{ij}\int d\boldsymbol{r} \nabla^\intercal x_i(t,\boldsymbol{r}) K_{ij} \nabla x_j(t,\boldsymbol{r})\,,
\end{align}
with $\nabla=(\partial_{r_1},\partial_{r_2})^\intercal$ being a two-dimensional gradient and the elements $U_{ij}$ being characteristic molecular interactions of the mean field~\cite{Flory1942, Huggins1941} and $K_{ij}$ corresponding to the energetic cost of forming the interfaces. Here, all thermodynamic entities are defined in the $RT$ unit, with $R$ being the gas constant and $T$ being the temperature. Furthermore, we assume that the matrices $\mathbb{U}$ and $\mathbb{K}$ are symmetric and have the following form:
\begin{align}
    \mathbb{U} = 
    \begin{pmatrix}
        0 & \chi \\
        \chi & 0
    \end{pmatrix}\,,
    \quad
    \mathbb{K} = 
    \begin{pmatrix}
        \kappa_1 & \kappa_2 \\
        \kappa_2 & \kappa_1
    \end{pmatrix}\,,
\end{align}
where $\chi$ and $\kappa_i$ are the interaction parameters. 

From the Helmholtz free energy, we obtain the chemical potential through functional derivatives
\begin{align}
    \mu_i = \frac{\delta F[\boldsymbol{x}]}{\delta x_i}\,,
\end{align}
calculated for the internal species in~\cref{eq:crn-brusselator} as
\begin{subequations}
\label{eq:cp}
\begin{align}
    \mu_1 &= \mu_1^0 + \log x_1 + \chi x_2 - \kappa_1 \Delta x_1-  \kappa_2 \Delta x_2\,, \\
    \mu_2 &= \mu_2^0 + \log x_2 + \chi x_1 - \kappa_2 \Delta x_1 - \kappa_1 \Delta x_2\,,
\end{align}
\end{subequations}
where $\mu_1^0$ and $\mu_2^0$ are the standard chemical potentials of species $\ch{X_1}$ and $\ch{X_2}$, respectively; and $\Delta = \partial_{r_1}^2 + \partial_{r_2}^2$ is a two-dimensional Laplace operator.  

We now define the reaction fluxes in \cref{eq:crn-brusselator}. For simplicity, we consider the almost irreversible regime where the net reaction fluxes are dominated by the forward reaction fluxes and furthermore assume the Arrhenius-like form~\cite{Avanzini2021} as
\begin{align}
\label{eq:j-vector}
    \boldsymbol{j} = (s_1 a, s_2 e^{2\mu_1 +\mu_2}, s_3 b e^{\mu_1}, s_4 e^{\mu_1})^\intercal \,,
\end{align}
where $s_\rho$ is the kinetic rate, and $a$ and $b$ are the concentrations of the chemostatted species $\text{A}$ and $\text{B}$, respectively. In what follows, we set $s_\rho = 1$ for $\forall \rho$. Combining \cref{eq:j-vector} with dynamical density functional theory~\cite{te2020classical}, we arrive at the nonideal RD system
\begin{align}
\label{eq:RD}
    \partial_t x_i &= D_i \nabla (x_i \nabla\mu_i) + \mathcal{J}_i\,,
\end{align}
where $D_i$ are the diffusion coefficients of species $i$ and the fluxes $\mathcal{J}_i$ have the following form
\begin{subequations}
\begin{align}
    \mathcal{J}_1 &= j_1 + j_2 - j_3 - j_4 = a + z_1^2 z_2 - (b + 1) z_1\,,\\
    \mathcal{J}_2 &=-j_2 + j_3 = - z_1^2 z_2 + b  z_1\,,
\end{align}
\label{flux}
\end{subequations}
with $z_i = e^{\mu_i}$ being the activity (fugacity). 
\section{\label{theo}Linear stability analysis}
\begin{figure*}
   \centering
    \includegraphics[width=0.9\textwidth]{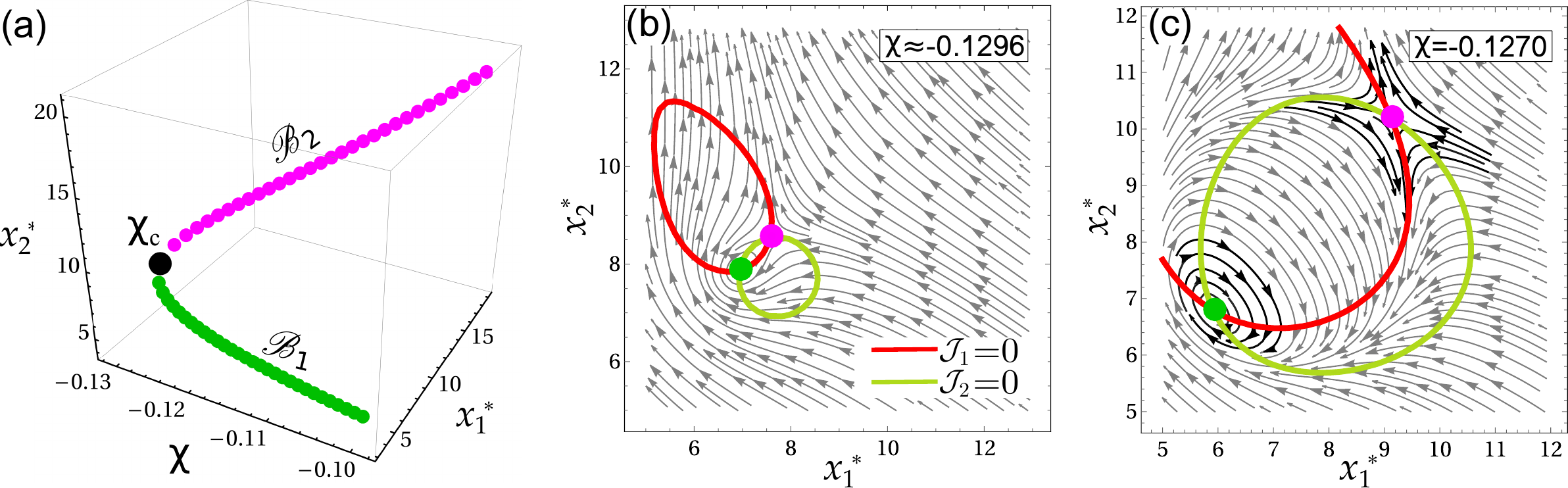}
    \caption{(a) The two branches ($\mathscr{B}_1$ and $\mathscr{B}_2$) of the homogeneous steady state for different interaction parameter $\chi$ that exist above the critical point $\chi_c \approx -0.1298$. Phase portrait of the homogeneous dynamics (\cref{eq:RD} without the diffusion term) for (b) $\chi \approx-0.1296$ and (c) $\chi=-0.1270$. The two steady state solutions are located at the intersections of the two nullclines, $\mathcal{J}_1=0$ and $\mathcal{J}_2=0$. They depart from each other from (b) to (c) as $\chi$ moves away from the critical point $ \chi_c$. The other parameters are $a=2.50$, $b=8$, $\mu_1^0=0$ and $\mu_2^0=0$.}
    \label{fig:criticalchi}
\end{figure*}

We notice the homogeneous steady state, $\mathcal{J}(x^*_1,x^*_2)=0$, in terms of activities reads
\begin{align}
\label{eq:ss-sol}
    \mathcal{J}_i(z_1^*,z_2^*) = 0 \rightarrow z_1^*= a\,, z_2^* = \frac{b}{a}\,,
\end{align}
where $z_i^* = e^{\mu_i^*}$ with $\mu_1^* = \mu_1^0 + \log x_1^* + \chi x_2^*$ and $\mu_2^* = \mu_2^0 + \log x_2^* + \chi x_1^*$.  We rewrite \cref{eq:ss-sol} in terms of $x_i^*$ as
\begin{subequations}
\label{eq:cp-ss}
\begin{align}
    a &= x_1^*e^{(\mu_1^0 + \chi x_2^*)}\,, \\
    \frac{b}{a} &= x_2^*e^{(\mu_2^0 + \chi x_1^*)}\,,
\end{align}
\end{subequations}
which implies a system of transcendental equations for $x_i^*$. For the ideal system, it gives the well-known fixed point $\boldsymbol{x}^*=(a, b/a)^\intercal$. However, in general, it can only be solved numerically. In~\cref{fig:criticalchi}, we plot the steady state solutions for different values $\chi<0$. There is a critical point $\chi_c$, such that for $\chi < \chi_c$, the steady state solution does not exist, and for $\chi > \chi_c$, it exhibits two branches.

We now study the stability of the homogeneous steady state of \cref{eq:RD}. 
Inserting $x_i = x_i^*+\delta x_i$, $\mu_i = \mu_i^* + \delta z_i/z_i^*$, and $\mathcal{J}_i = \delta \mathcal{J}_i = \sum_{k=1}^2\partial_{z_k}\mathcal{J}_i\delta z_k$ in \cref{eq:RD} while keeping only the linear terms in $\delta x_i$ and $\delta z_i$, we arrive at 
\begin{align}
\label{eq:stability-1}
    \partial_t \delta x_i = \sum_{k=1}^2\Big(\frac{D_ix_i^*}{z_i^*}\delta_{ik}\Delta + \frac{\partial \mathcal{J}_i}{\partial z_k}\Big)\delta z_k\,,
\end{align}
where $\delta_{ik}$ is a Kronecker symbol. After Fourier transforming \cref{eq:stability-1}, using the convention $\hat{f}(\boldsymbol{q}) = \int d\boldsymbol{r} \exp{(i \boldsymbol{r}\cdot \boldsymbol{q})}f(\boldsymbol{r})$, we arrive at
\begin{align}
    \partial_t \delta \hat{x}_i = \sum_{k=1}^2\Big(-\frac{D_ix_i^*\theta}{z_i^*}\delta_{ik} + \frac{\partial \mathcal{J}_i}{\partial z_k}\Big)\delta \hat{z}_k\,,
    \label{FT}
\end{align}
where $\theta \equiv q^2$. We note that the linear deviations $\delta \hat{z}_k$ and $\delta \hat{x}_k$ are related via the matrix
\begin{align}
\label{eq:matM}
    \mathbb{M} &= \Big[\frac{ \delta \hat{z}_k}{ \delta \hat{x}_m}\Big]_{\{k, m\}}=\Big[z^*_k\frac{\delta\hat{\mu}_k}{\delta\hat{x}_m}\Big]_{\{k, m\}} = \mathbb{Z}\cdot(\mathbb{M}_1 + \theta\mathbb{M}_2)\,,
\end{align}
written in terms of steady-state fugacities, 
\begin{align}
    \mathbb{Z} =
    \begin{pmatrix}
        a & 0\\
        0 & b/a
    \end{pmatrix}
    \,,\,
\end{align}
and interaction matrices,
\begin{align}
\label{eq:matZ-matM1-matM2}
    \mathbb{M}_1 = 
    \begin{pmatrix}
        (x_1^*)^{-1} & \chi \\
        \chi & (x_2^*)^{-1}
    \end{pmatrix}\,,\,
    \mathbb{M}_2 = 
    \begin{pmatrix}
        \kappa_1  & \kappa_2 \\
        \kappa_2  & \kappa_1
    \end{pmatrix}\,.
\end{align}
In matrix form,~\cref{FT} reads
\begin{align}
    \partial_t \delta \hat{\boldsymbol{x}} = (-\mathbb{D} \theta + \mathbb{G})\cdot\mathbb{M}\delta\hat{\boldsymbol{x}}\,,
\end{align}
where
\begin{subequations}
\label{eq:matD-matG}
\begin{align}
    \mathbb{D} &= \begin{pmatrix}
        D_1 x_1^*/z_1^* & 0 \\
        0 & D_2 x_2^*/z_2^*
    \end{pmatrix}\,,\\
\intertext{and}
    \mathbb{G} &=\Big[\frac{\partial \mathcal{J}_i}{\partial z_k}\Big]_{\{i, k\}}= \begin{pmatrix}
        b - 1 & a^2 \\
        -b & - a^2
    \end{pmatrix}\,.
\end{align}
\end{subequations}

The stability of \cref{eq:RD} is characterized by the eigenvalues $\lambda_+$ and $\lambda_-$ of the Jacobian matrix, 
\begin{align}
   \mathbb{J}
   &=(-\mathbb{D} \theta + \mathbb{G})\cdot\mathbb{Z}\cdot(\mathbb{M}_1 + \theta\mathbb{M}_2)\,.
   \label{eq:matJ}
\end{align}
Since the Jacobian matrix $\mathbb{J}$ is a $2\times2$ matrix with real entries, both eigenvalues must be either real or complex conjugates. This property is crucial for further analysis. We notice that this property does not hold in the presence of additional non-reacting species as in the Brusselator models\cite{aslyamov2023nonideal,cavalleriself}.

\begin{figure}
   \centering
    \includegraphics[width=\linewidth]{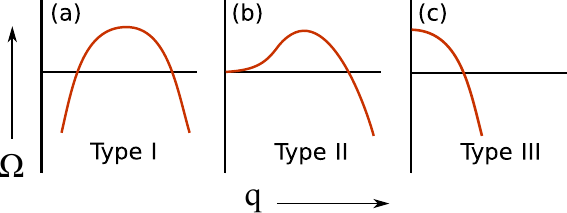}
    \caption{A schematic representation of dispersion curves, $\Omega(q)$, corresponding to three types of instabilities: (a) type I, (b) type II, and (c) type III. Here, $\Omega$ denotes the eigenvalue with the largest real part, and $q \equiv \sqrt{\theta}$ is the wavenumber.}
    \label{fig:types}
\end{figure}

Instabilities arise when the eigenvalue with the largest real part, which we denote $\Omega(\theta) = \text{Re}\,\lambda_+(\theta)$, becomes positive. More precisely, the system is unstable if
$\theta_{\text{max}}\equiv\arg\max_{\theta}\Omega(\theta) \geq 0 $ with $\Omega_\text{max} \equiv \Omega(\theta_\text{max}) > 0$. 
Under the Cross-Hohenberg classification \cite{Cross1993,cross2009pattern}, there are three types of instabilities distinguished by the behavior of $\Omega(\theta)$, as shown in \cref{fig:types}. Furthermore, following \cite{Cross1993} we consider the oscillation instabilities with $\text{Im}\,\lambda_n(\theta_\text{max})\neq 0$ denoting them as I-o, II-o, III-o and stationary instabilities with $\text{Im}\,\lambda_n(\theta_\text{max}) = 0$ denoted as I-s, II-s, III-s. 
We note that type~I-s and type~II-s instabilities are characterized by a nonzero inverse length scale, $\theta_{\text{max}} \neq 0$, which suggests the existence of structural patterns in the steady state~\cite{cross2009pattern}.

Before analyzing the instabilities, it is essential to restrict the parametric space to avoid ill-defined regimes. 
The system must have two real negative eigenvalues in the limit $\theta\to\infty$, where the Jacobian matrix becomes
\begin{align}
    \lim_{\theta \to \infty}\mathbb{J} = -\theta\mathbb{D}\cdot\mathbb{Z}\cdot
    \begin{pmatrix}
        (x_1^*)^{-1}+\kappa_1 \theta & \chi + \kappa_2 \theta\\
        \chi + \kappa_2 \theta & (x_2^*)^{-1}+\kappa_1 \theta
    \end{pmatrix}\,.
\end{align}
Thus, it must have a positive determinant, which implies 
\begin{align}
\label{chi_condition}
   (x_1^*x_2^*)^{-1} - (\chi)^2 < 0\rightarrow
    |\chi| < (x^*_1 x^*_2)^{-1/2}\,,\,\,\text{if}\,\,\,\kappa_{1,2} = 0 \;,
\end{align}
and otherwise
\begin{align}
\label{eq:kappa-condition-1}
   \kappa_1^2 - \kappa_2^2 > 0\rightarrow
    |\kappa_1|>|\kappa_2|\,,
\end{align}
as well as a negative trace that demands
\begin{align}
\label{eq:kappa-condition-2}
    \lim_{\theta\to\infty}\text{tr}\,\mathbb{J}<0\rightarrow\kappa_1 \geq 0 \,.
\end{align}

In our case, the eigenvalues are given by
\begin{align}
    \lambda_\pm = \frac{\operatorname{tr}\mathbb{J} \pm \sqrt{(\operatorname{tr}\mathbb{J})^2 - 4 \det \mathbb{J}}}{2}\,,
\end{align}
where $\text{Re}\lambda_+ \geq \text{Re}\lambda_-$ and
\begin{align}
\tr\,\mathbb{J} = \gamma + \beta\theta + \alpha \theta^2\,,
    \label{eq:tr}
\end{align}
with
\begin{subequations}
\label{eq:tr-terms}
\begin{align}
\label{eq:gamma}
    \gamma &= a[(b-1)/x_1^* - b/x_2^*]\,,\\
\label{eq:beta}
    \beta &= -(D_1 + D_2 + a \kappa_1) < 0\,,\\
\label{eq:alpha}
    \alpha & = -\kappa_1(D_1  x_1^* + D_2 x_2^*) \leq 0 \;,
\end{align}
\end{subequations}
where we used~\cref{eq:kappa-condition-2} for inequalities. For the ideal limit, we write $\alpha^\text{id} = 0$, $\beta^\text{id} = -(D_1+D_2)<0$, 
$\gamma^\text{id} = (b-1)e^{\mu_1^0} - a^2e^{\mu_2^0}$. We notice that $\operatorname{tr}\mathbb{J}$ and $\det \mathbb{J}$ are quadratic in $\theta$.

\section{\label{sec:ar}Classification of instabilities}
We now analyze the Cross–Hohenberg types of instabilities that can arise in our system. We begin with an analytical derivation and then present numerical simulations for systems with competing instabilities. The classification is presented in \cref{tab:types}. 

\subsection{Complex conjugate eigenvalues: Oscillations}
We start with complex eigenvalues that indicate oscillatory behavior. Since the complex eigenvalues $\lambda_+$ and $\lambda_-$ share the same real part, 
\begin{align}
    2\Omega = \lambda_+ + \lambda_- = \tr\,\mathbb{J}(\theta) = \gamma + \beta\theta + \alpha \theta^2\,,
    \label{realeigen}
\end{align}
where we used \cref{eq:tr}.

\subsubsection{Type I-o and II-o are impossible.}

The curves $\Omega(\theta)$ in \cref{fig:types}(a) and (b) exhibit a maximum at a finite wavenumber $\theta_\text{max}>0$. This requires that the derivative $\partial_\theta \Omega(\theta)$ changes from positive to negative while crossing $\theta_\text{max}>0$. However, this is impossible since $\partial_\theta \Omega(\theta)=\beta/2+\alpha\theta \leq 0$. Thus, oscillations of types I-o and II-o are impossible.

\subsubsection{Type III-o is possible.}
The curve $\Omega(\theta)$ in \cref{fig:types}(c) has a positive maximum at $\theta=0$ and then decreases monotonically, $\partial_\theta \Omega(\theta) < 0$. This is possible for $\Omega(0) = \gamma/2 > 0$. We show the numerical results for type III-o in \cref{fig:PANEL1}.
In the ideal case, we have $\Omega^\text{id} = \gamma^\text{id} - (D_1 + D_2) q^2$, which allows the type III-o instability for $\gamma^\text{id} > 0$.
\subsection{\label{typereal}Real eigenvalues: Stationary patterns}
We now proceed to analyze the case of two real eigenvalues, where $\Omega = \text{max}(\lambda_+,\lambda_-)$. 

\subsubsection{Type I-s is possible}
\begin{figure*}
   \centering
    \includegraphics[width=\textwidth]{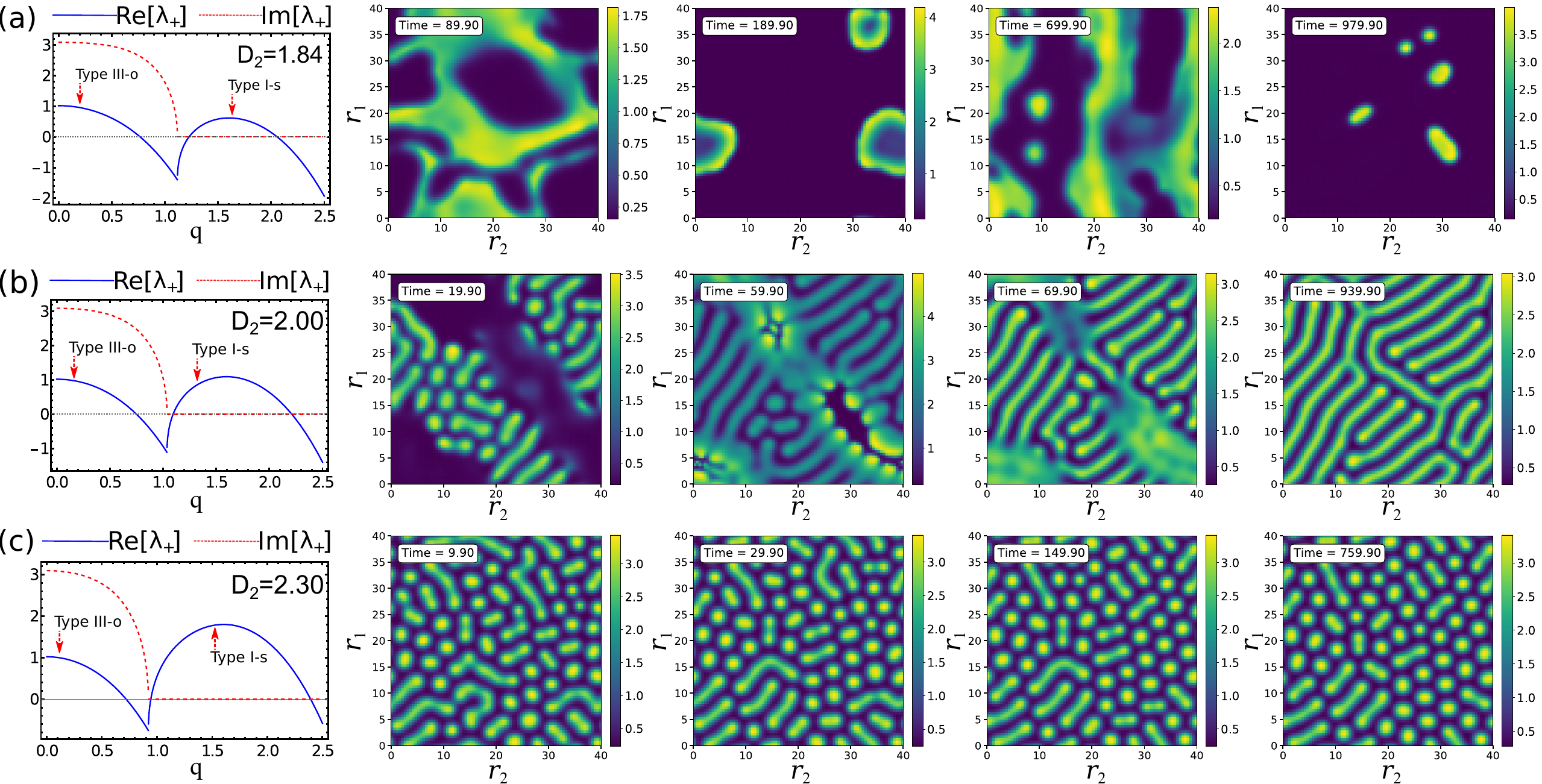}
    \caption{~Dispersion curves, and snapshots of $x_1$ concentration field at different time points for three different inhibitors' diffusion coefficients: (a) $D_2=1.84$, dominating type III-o instability, (b) $D_2=2$, comparable type I-s and III-o, (c) $D_2=2.30$, stronger type I-s instability. Stationary spatial patterns emerge in (b) and (c). The remaining simulation parameter values are $x_1^*=1.72$, $x_2^*=2.47$, $a=2.50$, $b=8$, $\mu_1^0=0$, $\mu_2^0=0$, $D_1=1$, $\chi=0.15$, $\kappa_1=0.10$, and $\kappa_2=0.05$.}
    \label{fig:PANEL1}
\end{figure*}

At the critical point $\theta_c$, the curve $\Omega(\theta)$ touches the $\theta$-axis,
\begin{align}
\label{eq:real-ev-condition-1}
    \Omega_c =\Omega(\theta_c) = 0 \,,\quad \partial_\theta \Omega_c = 0\,.
\end{align}
In terms of eigenvalues, this equivalently means
\begin{align}
\label{eq:real-ev-condition-2}
    &\det\mathbb{J}(\theta_c)=0\,,\quad\partial_\theta\det\mathbb{J}(\theta_c)=0\,.
\end{align}
Using~\cref{eq:matJ}, we find two scenarios. The \emph{first scenario} reads
\begin{subequations}
\label{eq:real-type-1}
    \begin{align}
    \label{eq:real-type-1-det=0}
    \det(-\mathbb{D}\theta_c+\mathbb{G}) &= 0\,, \\
    \partial_\theta\det(-\mathbb{D}\theta_c+\mathbb{G}) &= 0 \,.
\end{align}
\end{subequations}
Since the determinants in \cref{eq:real-type-1} are quadratic polynomials in $\theta_c$, \cref{eq:real-type-1} implies the existence of a unique, positive root $\theta_c$. It is equivalent to setting the discriminants of the quadratic equation $\det(-\mathbb{D}\theta_c + \mathbb{G})=0$ to zero, that results in 
\begin{align}
\label{eq:real-type-1-D-ratio}
    \frac{D_1}{D_2} &= \frac{(\sqrt{b} \pm 1)^2 x_2^*}{b x_1^*}\,.
\end{align} 
The corresponding unique root of \cref{eq:real-type-1-det=0} reads
\begin{align}
\label{eq:real-type-1-lc}
    \theta_c &=  \frac{ab}{D_2 x_2^* (\sqrt{b} \pm 1)}\,.
\end{align}
Thus, the type~I-s instability is possible, giving rise to stationary patterns, as indicated by the nonzero critical inverse length scale $\theta_c \neq 0$ (see \cref{fig:PANEL1}).

We notice that \cref{eq:real-type-1-D-ratio} is an explicit expression for the ratio $D_1 / D_2$, since the fixed point $\boldsymbol{x}^*$ does not depend on the diffusion coefficients $D_i$. Therefore \cref{eq:real-type-1-D-ratio} may be satisfied for arbitrary $\boldsymbol{x}^*$ and arbitrary values of the interaction parameter $\chi$.
Accordingly, the type~I-s instability remains possible in the ideal limit. We therefore conclude that interactions do \emph{not} explicitly generate the instability in this setting; rather, they may exert implicit control through their influence on the homogeneous steady state $\boldsymbol{x}^*$.

The \emph{second scenario} has an explicit interaction origin and could be written as
\begin{subequations}
\label{eq:real-type-2}
    \begin{align}
    \det(\mathbb{M}_1+\theta_c \mathbb{M}_2) &= 0\,, \\
    \partial_\theta \det(\mathbb{M}_1+\theta_c \mathbb{M}_2) &= 0\,,
\end{align}
\end{subequations}
which corresponds to the zero discriminant of $\det(\mathbb{M}_1 + \theta_c \mathbb{M}_2)=0$ and results in 
\begin{align}
\label{eq:real-type-2-k1}
 \frac{\kappa_1}{\kappa_2} &=\frac{2x_1^*x_2^*[\chi(x_1^*+x_2^*)\pm\sqrt{(x_1^*-x_2^*)^2(\chi^2-(x_1^*x_2^*)^{-1})}]}{(x_2^*-x_1^*)^2+(2\chi x_2^* x_1^*)^2} \,,
%  \label{eq:real-type-2-theta_c}
% \theta_c^2 &=\frac{(x_1^*x_2^*)^{-1}-\chi^2}{(\kappa_1^2-\kappa_2^2)}\,,
\end{align}
where $x_i^*$ does not depend on $\kappa_i$.
Since $\kappa_i$ are real, we must have $\chi^2-(x_1^*x_2^*)^{-1}\geq 0$ from the square root in \cref{eq:real-type-2-k1}. Combining this condition with $\kappa_1^2> \kappa_2^2$ [\cref{eq:kappa-condition-1}], we find that the only possible condition is $(x_1^*x_2^*)^{-1}=\chi^2$, which results in zero critical wavenumber $\theta_c=0$. Thus, a type I-s with the second (pure interactive) scenario does not exist. 
\begin{figure}
   \centering
    \includegraphics[width=0.8\linewidth]{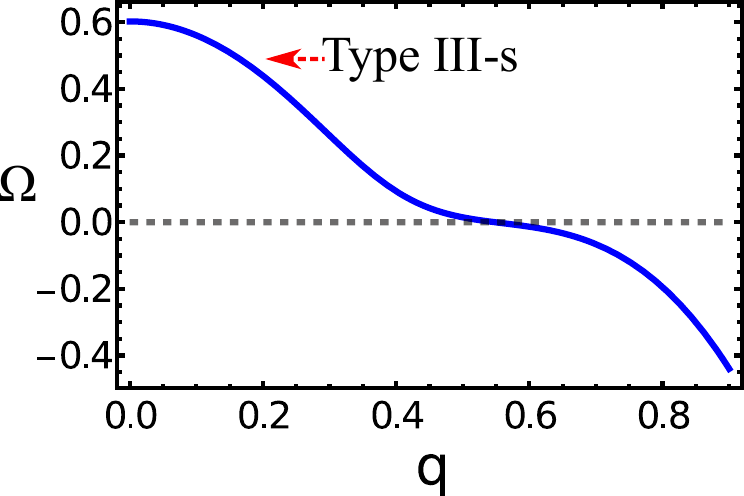}
    \caption{Type III-s instability in the dispersion curve. Parameters are specified as $a=2.50$, $b=8$, $\chi= -0.11$, $D_1=1$, $D_2=2$, $\kappa_1=0.1$, and $\kappa_2=0.05$.}
    \label{fig.typeiii}
\end{figure}
\subsubsection{Type II-s is impossible}
Type II-s implies that as $\theta$ approaches zero, $\Omega(\theta) > 0$ and eventually $\Omega(0) = 0$, while the second eigenvalue remains negative. 
This means that as $\theta$ approaches zero, $\det \mathbb{J}(\theta)<0$ tends to zero.
Using \cref{eq:matJ}, we find that
\begin{align}
\label{eq:detJ0}
\det \mathbb{J}(0) &=a^2b\big[(x_1^*x_2^*)^{-1}-\chi^2\big] = 0
\end{align}
when $\chi = \pm (x_1^*x_2^*)^{-1/2}$. By expanding $\det \mathbb{J}(\theta)$, we get
\begin{align}
\label{eq:detJ-near-zero}
     \det \mathbb{J}(\theta) = \frac{a^2 b}{x_1^* x_2^*}\Big[\mp 2\kappa_2\sqrt{x_1^* x_2^*}+\kappa_1(x_1^*+x_2^*)\Big]\theta + \mathcal{O}(\theta^2)\,.
\end{align}
Since, using \cref{eq:kappa-condition-1} and \cref{eq:kappa-condition-2},
\begin{align}
\label{eq:detJ-bound}
    \mp 2\kappa_2\sqrt{x_1^* x_2^*}+\kappa_1(x_1^*+x_2^*)&\geq\nonumber\\
    -2|\kappa_2|\sqrt{x_1^* x_2^*}+\kappa_1(x_1^*+x_2^*)&>-2\kappa_1\sqrt{x_1^* x_2^*}+\kappa_1(x_1^*+x_2^*)\nonumber\\
    &=\kappa_1\Big(\sqrt{x_1^*}-\sqrt{x_2^*}\Big)^2 \geq 0\,,
\end{align}
we prove that $\det\mathbb{J}(\theta)>0$ as $\theta$ approaches zero. Therefore, type II-s is impossible since it requires $\det\mathbb{J}(\theta)\to -0$ for $\theta\to +0$.  

\begin{table}
\caption{\label{tab:types}Types of instabilities that can arise in the nonideal Brusselator RD framework. Stationary and oscillatory instabilities are denoted by \textbf{s} and \textbf{o}, respectively.}
\begin{ruledtabular}
\begin{tabular}{cccccc}
\multicolumn{2}{c}{\textbf{Type I}} & 
\multicolumn{2}{c}{\textbf{Type II}} & 
\multicolumn{2}{c}{\textbf{Type III}} \\
\hline
\textbf{s} & \textbf{o} & \textbf{s} & \textbf{o} & \textbf{s} & \textbf{o} \\
\midrule
\textcolor{green}{\checkmark} & \textcolor{red}{\(\times\)} & 
\textcolor{red}{\(\times\)} & \textcolor{red}{\(\times\)} & 
\textcolor{green}{\checkmark} & \textcolor{green}{\checkmark} \\
\end{tabular}
\end{ruledtabular}
\end{table}
\begin{figure}
   \centering
    \includegraphics[width=0.8\linewidth]{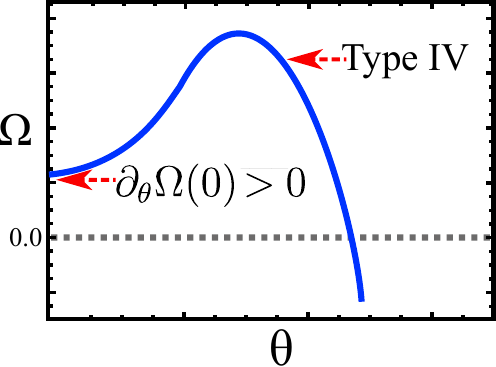}
    \caption{The sketch of the type IV dispersion curve.}
    \label{fig:tiv}
\end{figure}
\begin{figure*}
   \centering
    \includegraphics[width=\textwidth]{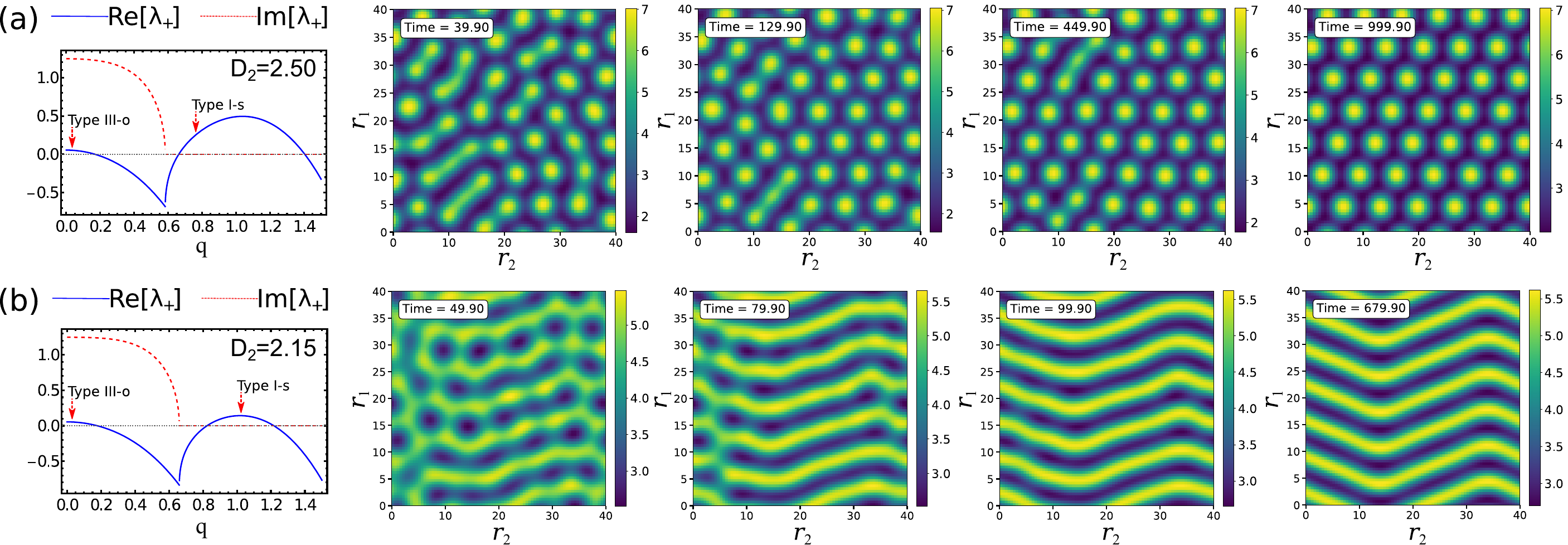}
    \caption{~Dispersion curves and corresponding snapshots of the $x_1$ concentration field at different time points for attractive interactions, shown for (a) $D_2=2.50$, and (b) $D_2=2.15$. Panel (a) exhibits droplet-like pattern, whereas panel (b) shows a zigzag pattern. %\textit{Inset:} Dispersion curves of the type III-o lie well above the neutral stability line. 
    Other parameters are: $a=2.50$, $b=8$, $D_1=1$, $\mu_1^0=0$, $\mu_2^0=0$, $\chi=-0.11$, $\kappa_1=0.10$, and $\kappa_2=0.05$.}
    \label{fig:PANEL2}
\end{figure*}

\subsubsection{Type III-s is possible}
Type III-s implies $\theta_\text{max}=0$ and  $\Omega(0) > 0$. From \cref{realeigen} and \cref{eq:gamma}, we find $\gamma=a\frac{(b-1)x_2^*-bx_1^*}{x_1^*x_2^*}>0$. 
An example is shown in~\cref{fig.typeiii}.
We notice that the same criterion holds for the type III-s instability in the ideal case. 

\subsection{Type IV instability}
In our analysis, we implicitly assumed that the eigenvalues are finite everywhere and that the real part of the eigenvalues changes smoothly around the maximum~\cite{cross2009pattern}. 
Under these assumptions, a new type of instability that is not part of the Cross-Hohenberg classification, type IV, could also be expected; see \cref{fig:tiv}. 
It requires a positive eigenvalue at $\theta=0$ (similar to type III), and a positive maximum at a finite wavenumber $\theta > 0$ (akin to types I and II). However, we show that type IV instability is impossible in our system for both real and complex eigenvalues. 
For complex eigenvalues, we have $\partial_\theta \Omega(0) = \beta <0$, which contradicts \cref{fig:tiv}. 
From \cref{eq:detJ-near-zero,eq:detJ-bound}, we write $\det \mathbb{J}(0) \geq 0$ and, 
for real eigenvalues, we calculate
\begin{align}
\label{eq:detJ0-Omega}
    \partial_\theta \det \mathbb{J}(0)  = \partial_\theta\Omega(0) \underbrace{(\lambda_-(0) - \Omega)}_{\leq 0} + \underbrace{\Omega \beta}_{< 0} \geq 0.
\end{align}
where we used \cref{realeigen} for $\partial_\theta \lambda_-(0) = \partial_\theta(\text{tr}~\mathbb{J} - \Omega)|_{\theta=0}$ and $\partial_\theta \text{tr}~\mathbb{J}(0) = \beta < 0$.
Thus, \cref{eq:detJ0-Omega} requires $\partial_\theta\Omega(0) < 0$.

\subsection{\label{sr}Numerical simulations}
We numerically solve the RD system described in~\cref{eq:RD} for a nonideal Brusselator model. As an initial condition, we use the homogeneous steady state $x^*_i$ with an additional Gaussian white noise of zero average and $(0.1x_i^*)^2$ variance.
Simulations are carried out on a finite two-dimensional square domain of size $L = 40$ with periodic boundary conditions in both spatial directions. All simulations are performed using the py-pde package~\cite{py-pde}. 

We simulate our system for two types of molecular interactions, repulsive ($\chi > 0$) and attractive ($\chi < 0$).
We find that the dispersion relations generically exhibit both features of type III-o and type I-s instabilities, regardless of the interaction type; see \cref{fig:PANEL1,fig:PANEL2}.
For type I, there is a nonzero fastest growing mode which is reflected in the structure of the corresponding stationary patterns. For type I-s dominance, we obtain labyrinthine, droplets, and zigzag stationary patterns in~\cref{fig:PANEL1,fig:PANEL2}. As analytically shown for the critical point [\cref{eq:real-type-1-lc}], the fastest mode depends on the diffusion coefficients. Moreover, varying the diffusion coefficient one could change the dominant instability and strongly modify the stationary pattern or even destroy it. We illustrate it in~\cref{fig:PANEL1}, for $\chi > 0$, increasing $D_2$ leads to a change in the dominant instability from type III-o to type I-s.

\section{\label{dis}Conclusion} 
In this study, we examined a thermodynamically consistent nonideal Brusselator model, through the lens of the Cross–Hohenberg classification scheme. Using linear stability analysis, we demonstrated that only type I and type III instabilities can emerge for real eigenvalues, while type III instability is the sole possibility for complex eigenvalues [\cref{tab:types}]. 
We derived the analytical conditions governing the feasibility of these instabilities.
Notably, we found that the energetic contribution alone is insufficient to induce type I instability and, hence, spatial pattern formation in the system. 
We notice that this restriction does not hold for the Brusselator RD system with an additional non-interacting species, as recently shown in \cite{cavalleriself}. 

For both repulsive and attractive interaction scenarios, we numerically captured distinct patterns that emerged in the simulation for different diffusion coefficient values, which correctly reflected the instability dominance scenarios encoded in the accompanying dispersion curves. Recently, a similar dominance scenario between static and oscillatory instabilities has been explored in a nonreciprocal
Model B~\cite{Sollich2025nonreciprocal}. The core autocatalytic reaction examined here has also been incorporated into a thermodynamically consistent model coupling chemical and mechanical forces that can give rise to self-propulsion~\cite{VossUwe2025}. 

Our analysis focused on a specific RD system, but the underlying methodology is applicable to a broader class of two-component RD models~\cite{schlogl1972chemical, GrayScott}. Similar analysis could also be extended to study systems involving three or more interacting internal species~\cite{threespecies, GolestanianCIPS, DZ2024, bauermann2025spatiotemporal}.
Our approach could also be extended to study modern energy devices (batteries and pseudocapacitors), where two-component reaction-diffusion systems (positive and negative ions) play an important role in the electrochemistry that combines electrostatic transport equations (such as Poisson-Nernst-Planck) \cite{aslyamov2022relation,aslyamov2022analytical} with the kinetics of the redox reaction on the surface of the electrode \cite{bazant2017thermodynamic,aslyamov2025faradaic}. Extensions incorporating fluctuations could also be considered to study their effect on the phase separation phenomena \cite{Tiani2023, MasiRevModPhys, yanagisawa2025phase}.

\begin{acknowledgments}
P.K. and M.E. acknowledge the financial support from
projects C21/MS/16356329/CHEMCOMPLEX and C24/MS/18933049/NEQPHASETRANS funded by the Fonds National de la Recherche-FNR, Luxembourg;  
T.A. acknowledges the financial support from
project ThermoElectroChem (C23/MS/18060819) funded by the Fonds National de la Recherche-FNR, Luxembourg. 

\end{acknowledgments}
\bibliography{biblio}
\end{document}